\documentclass[aps,twocolumn,prl,color,psfig,epsf]{revtex4}
\usepackage{amsmath}
\usepackage{color}
\usepackage{amsfonts}
\usepackage{epsf}
\usepackage{graphicx}
\usepackage{epstopdf}
\usepackage{ulem}
\usepackage[title]{appendix}

\begin{document}

\title{Mixtures of blue phase liquid crystal with simple liquids: \\ elastic emulsions and cubic fluid cylinders}

\author{J.~S. Lintuvuori$^{1}$, K. Stratford$^2$, M.~E. Cates$^{3}$, D. Marenduzzo$^4$}
\affiliation{$^1$ Univ. Bordeaux, CNRS, LOMA, UMR 5798, F-33405 Talence, France\\
  $^2$ EPCC, The University of Edinburgh, Edinburgh EH9 3FD, United Kingdom\\
  $^3$ DAMTP, Centre for Mathematical Sciences, University of Cambridge, Cambridge CB3 0WA, United Kingdom\\
  $^4$ SUPA, School of Physics and Astronomy, The University of Edinburgh, Edinburgh EH9 3FD, United Kingdom}

\begin{abstract}
We investigate numerically the behaviour of a phase-separating mixture of a blue phase I liquid crystal with an isotropic fluid. The resulting morphology is primarily controlled by an inverse capillary number, $\chi$, setting the balance between interfacial and elastic forces. When $\chi$ and the concentration of the isotropic component are both low, the blue phase disclination lattice templates a cubic array of fluid cylinders. For larger $\chi$, the isotropic phase arranges primarily into liquid emulsion droplets which coarsen very slowly, rewiring the blue phase disclination lines into an amorphous elastic network. Our blue phase/simple fluid composites can be externally manipulated:  an electric field can trigger a morphological transition between cubic fluid cylinder phases with different topologies.
\end{abstract}

\maketitle

Composite materials based on complex fluids, such as liquid crystalline droplets or emulsions~\cite{lcdroplet,lcdroplet2,EtienneDroplet,lcemulsion,lcemulsion2} and colloid-liquid crystal mixtures~\cite{miha1,miha2,tiffany,natcom}, have received a lot of attention of late. This is due both to their rich and often unexpected physical behaviour, and to their potential as soft materials with unusual optical, flow and mechanical properties. These features may lead to technological applications. For example, cholesteric droplets can be utilised as tunable microlasers~\cite{microlaser} or in optofluidics~\cite{EtienneChiralSorting,EtienneChiralTrapping}. Likewise, photopolymerised blue phase disclination networks make intriguing switchable electro-optic devices~\cite{Flyn}.

From a fundamental viewpoint, the richness in behaviour of many such composites derives from the competition between different length scales (e.g., droplet size, cholesteric pitch, defect size), and from the contest between interfacial and elastic energies~\cite{colloidlcreview,colloidlcreview2,juhoanneinterface}. The free energy landscape is often glassy and possesses multiple metastable structures, where energy barriers dwarf thermal energies. Structural arrest in this landscape may pave the way to energy-saving multistable devices which retain memory of their state in the absence of an external (electric or magnetic) field~\cite{bistable}. 

Here, we study computationally a binary mixture of a blue phase (BP) liquid crystal and an isotropic liquid, and characterise the resulting behaviour and dynamics. BPs are remarkable liquid crystals (LCs) displaying a 3D network of disclination lines~\cite{mermin}. Without external fields, BP disclinations may have either cubic symmetry (blue phases I and II, BPI and BPII), or be amorphous (blue phase III~\cite{bp3}). The typical length scale of the network is close to the pitch of the cholesterics which form the BP, normally a few hundred nm. Dispersing colloidal nanoparticles or polymers inside BPs leads to a dramatic increase in their range of thermodynamic stability, as the energetically costly disclinations are covered by these inclusions~\cite{BPpolymer,miha2}. 

Our simulations reveal a number of striking physical properties for composites of BPI with simple liquids. If the interfacial tension between the two, $\sigma$, is sufficiently small, then the isotropic component arranges into very long liquid tubes surrounding the BP disclinations, creating what we call a ``cubic fluid cylinder phase'' -- this is similar to the emulsified blue phase first theorised in~\cite{bpemul}, although in our case no surfactants are required.
Just as polymers or nanoparticles, such liquid tubes can stabilise the network thermodynamically. In turn they are stabilized against the Rayleigh instability by the BP elasticity. For larger $\sigma$, the phenomenology is completely different: there is a transition to a regime where quasi-spherical isotropic droplets grow within the BP matrix. Coarsening does not proceed indefinitely, but is arrested before full phase separation is reached to create an amorphous, elastic emulsion. This transition can be understood via a simple mean field theory which identifies an elastic capillary number as the key control parameter. Finally, we show that these composites are reconfigurable: the topology of the cubic fluid cylinder phase can be altered by an electric field.

{\it Simulation methods}: To simulate a BP mixture we consider a free energy functional ${\cal F}=\int f dV$. Its density consists of two parts, $f=f_{\phi} + f_{Q}$, describing local mixture composition and LC ordering respectively. The first part can be written in terms of a compositional order parameter field $\phi$ as
\begin{equation}
 f_{\phi}=-\frac{a}{2}\phi^2+\frac{b}{4}\phi^4+\frac{\kappa}{2}\left(\nabla \phi\right)^2
\end{equation}
where {{$a=b$}} and $\kappa$ are positive constants. For a cholesteric LC, $f_{Q}$ can be expressed in terms of a traceless and symmetric order parameter tensor ${\bf Q}$,
\begin{align}
\label{eq:FreeEnergy}
{f_{Q}} & = \tfrac{A_0}{2} \bigl( 1 - \tfrac{\gamma(\phi)}{3} \bigr) Q_{\alpha \beta}^2 
           - \tfrac{A_0 \gamma(\phi)}{3} Q_{\alpha \beta}Q_{\beta \gamma}Q_{\gamma \alpha} \notag \\
         &  + \tfrac {A_0 \gamma(\phi)}{4} (Q_{\alpha \beta}^2)^2  - \frac{\epsilon_a}{12\pi}E_{\alpha}Q_{\alpha\beta}E_{\beta}\\
           &+ \tfrac{K}{2}\bigl( \nabla_{\beta}Q_{\alpha \beta}\bigr)^2 
           + \tfrac{K}{2} 
           \bigl( \epsilon_{\alpha \gamma \delta} \nabla_{\gamma} Q_{\delta \beta} 
           + 2q_0 Q_{\alpha \beta} \bigr)^2, \notag
\end{align}
where repeated Greek indices (denoting Cartesian components) are summed. In Eq.~\ref{eq:FreeEnergy}, $A_0$ gives the energy scale, $K$ is the elastic constant and $q_0=2\pi/p$ fixes the equilibrium pitch length $p$. We include an electric field $\mathbf{E}$ (with associated dielectric constant anisotropy $\epsilon_a$), which will be considered at the end of this Letter. The quantity $\gamma(\phi) = \gamma_0 + \delta(1+\phi)$ controls the order. We choose $\gamma_0$ and $\delta$ such that $\gamma(-1)$ is below the threshold for isotropic-cholesteric transition, whereas $\gamma(+1)$ is above it. {{The full parameter list is given in the Appendix together with details of initial conditions, and a mapping between simulation and physical units in~\cite{Mapping}. We highlight that the chirality, $\kappa_{LC}=\sqrt{108 Kq^2_0/(A_0\gamma)}$, and the reduced temperature, $\tau=27(1-\gamma/3)/\gamma$ were chosen as $\kappa_{LC}\approx 0.7$ and $\tau\approx -0.25$ throughout, to favour BPI thermodynamically~\cite{OliBP}.}}
We do not include an explicit surface anchoring term; we expect its addition to further enrich the possibilities for composite design~\cite{natcom}.

The evolution of the LC order ${\bf Q}$ is given by the Beris-Edwards equation~\cite{beris} 
\begin{equation}
\left(\partial_t+u_\alpha\partial_\alpha\right){\bf Q}-{\bf S}\left({\bf u},{\bf Q}\right)=\Gamma{\bf H},
\end{equation}
where ${\bf S}\left({\bf u},{\bf Q}\right)$ accounts for flow-induced LC rotation (see Appendix), $\Gamma$ is related to the rotational viscosity and the molecular field ${\bf H} = - \left[\frac{\delta {\cal F}}{\delta {\bf Q}} - \frac{1}{3}{\bf I}\mathrm{Tr} \frac{\delta {\cal F}}{\delta {\bf Q}}\right]$. The velocity field~$u_{\alpha}$ satisfies a Navier-Stokes equation where the stress tensor includes elastic contributions~{{\cite{hybridLB}} (see Appendix). 
The equation of motion for $\phi$ is 
\begin{equation}
 \partial_t\phi+\partial_\alpha\left(\phi u_\alpha\right) = M\nabla^2\mu
\end{equation}
where $M$ is a constant mobility and $\mu=\frac{\delta {\cal F}}{\delta\phi}$ is the chemical potential. We solve the equations of motion using a hybrid lattice Boltzmann algorithm~\cite{hybridLB}. To minimize discretization artefacts we include thermal noise in the Navier-Stokes equation~\cite{ThermalLB}; this is set at a level too weak to change qualitatively the structural dynamics on the length scales of interest here~\cite{notetemp}. Simulations were carried out in a $128\times 128\times128$ cubic periodic simulation box, {{where the BPI unit cell size $\lambda=p/\sqrt{2}$~\cite{OliBP} is either $32$ or $64$ lattice units -- the defect core $\xi$ and interfacial width $\xi_{\phi}$ are both much smaller, and of the order of a few lattice units (see Appendix)}}.
 
\begin{figure}
\includegraphics[width=\columnwidth]{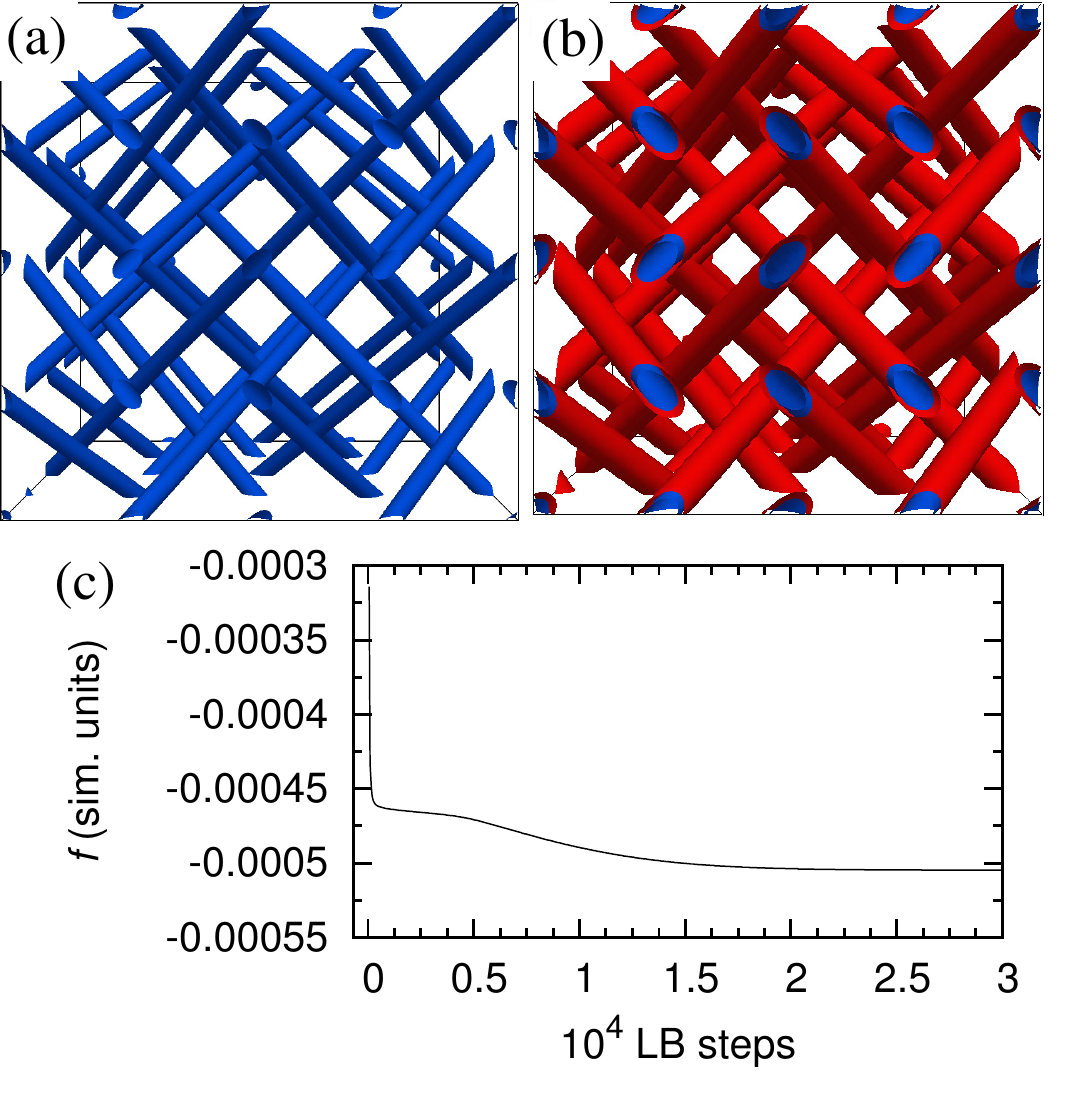}
\caption{Formation of a cubic fluid cylinder phase. (a) BPI disclination network (blue ribbons). (b) In a 90:10 BP:isotropic mixture with low $\sigma$ ($\chi \equiv \sigma \xi/K \approx 0.019$), the isotropic fluid self-organises to form tubes following the cubic symmetry of the BPI disclination network. (c) The free energy of the system as a function of time $t$. 
{{(Simulations performed using parameter set {\it A} (see Appendix).)}} 
}
\label{fig:BPIstruct}
\end{figure}

\begin{figure*}
\includegraphics[width=\textwidth]{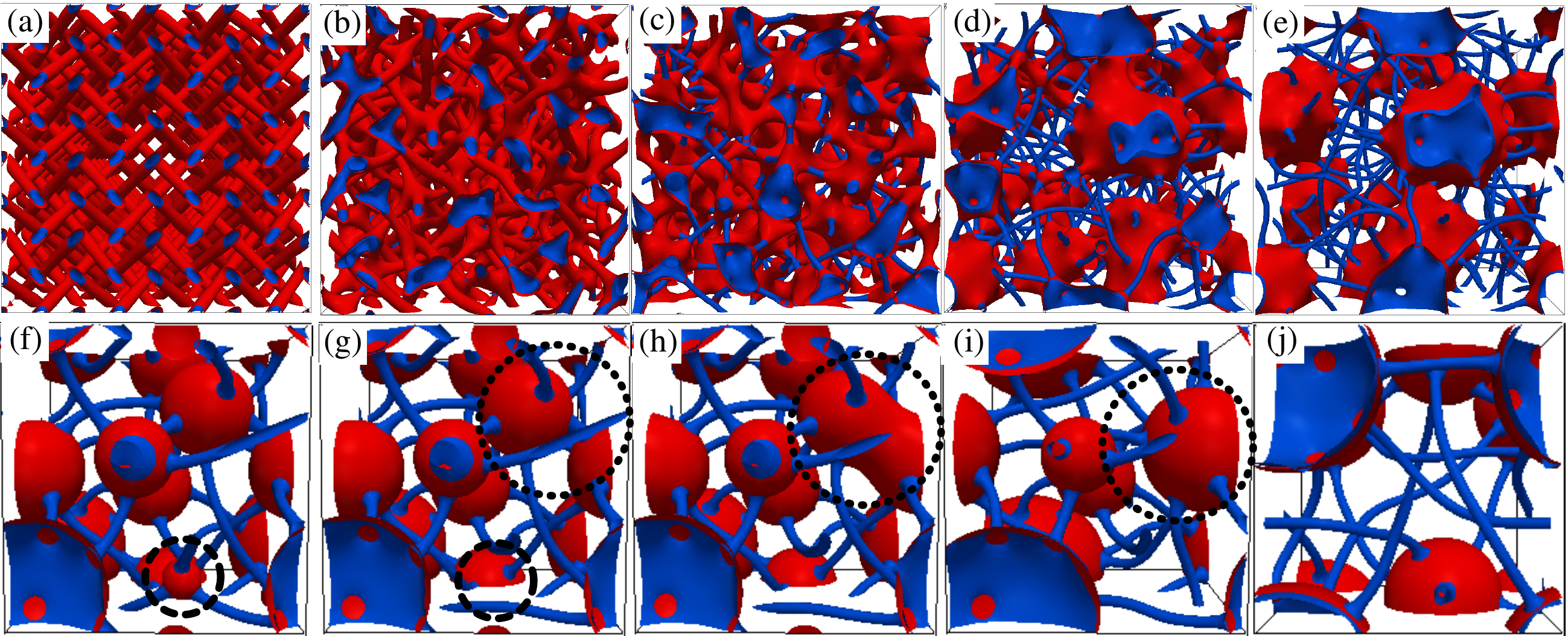}
\caption{Dynamics of BP mixtures with different values of $\chi$ and LC:isotropic composition. (a-e) Dynamics of a low interfacial tension system ($\chi\approx 0.019$), with 85:15 composition, starting from an initially mixed state. Snapshots correspond to: (a)~$t = 2\times 10^5$ (simulation steps); (b)~$t=7\times 10^5$, (c)~$t=1\times 10^6$, (d)~$t=2.5\times 10^6$, (e)~$t=9.6\times 10^6$. {{(Simulations performed using parameter set {\it B} (see Appendix).)}} 
(f-j) Dynamics of an 80:20 BP mixture with higher interfacial tension ($\chi\approx 1.9$). 
Snapshots correspond to: (f)~$t=1.8\times10^6$, (g)~$t=2.1\times 10^6$, (h)~$t=2.2\times 10^6$, (j)~$t=6.4\times 10^6$. 
{{(Simulations performed using parameter set {\it C} (see Appendix).)}}
}
\label{fig2}
\end{figure*}

{\it Results}: All the simulations presented have the isotropic component as the minority phase. Figure~\ref{fig:BPIstruct} shows the behaviour of a 90:10 LC:isotropic mixture with a low value of the interfacial tension ($\sigma=5.4\times 10^{-4}$ in simulation units, or $\sim 10^{-5}$Nm$^{-1}$ in physical units~\cite{Mapping}). The components rapidly demix so that the the isotropic fluid is templated by the cubic symmetry of the emerging BPI disclination pattern (Suppl.~Movies~1 {{and 2}}), in agreement with theoretical predictions~\cite{bpemul}. 
We name the resulting structure a cubic fluid cylinder phase, because the topology of the isotropic component is templated by that of the cubic BPI. The elastic character of the latter then stabilizes the fluid cylinders against the Rayleigh instability that would cause them to break into droplets if surrounded instead by a second fluid.

To explain why a cubic fluid cylinder phase forms, we propose a simple mean-field argument. The elastic free energy cost per unit length $E_{\rm el}$ associated with a cylinder of radius $r$ containing a straight disclination line (with charge $s=-1/2$, relevant for BPs~\cite{mermin}) of core radius $\xi$ can be estimated by assuming that the defect core is isotropic, as follows,
\begin{eqnarray}\label{elasticenergy}
E_{\rm el} & = & E_0(r/\xi)^2 \qquad \qquad \qquad \, \, \, \, {\rm for} \, \, r<\xi \\ \nonumber
E_{\rm el} & = & \frac{\pi K}{4} \log(r/\xi)+E_0 \qquad {\rm for} \, \, r\ge\xi.
\end{eqnarray}
In Eq.~\ref{elasticenergy}, $E_0/\xi^2\sim A_0$ is the uniform free energy density associated with the melting of the isotropic defect core. (In our simulations, and in reality, the core is not fully melted but weakly biaxial; this would change only prefactors above.) This needs to be compared with the interfacial free energy per unit length associated with the formation of a cylindrical tube of isotropic fluid of radius $r$, $E_{\rm in}=2\pi \sigma r$. The fluid tube forms if $\delta E=E_{\rm in}-E_{\rm el}<0$, in which case the selected radius, $r^*$, is found by minimising $\delta E$, giving $r^*=K/(8\sigma)$. This solution only holds for $\sigma<\sigma^{*}$, where $\sigma^*$ is given by the formula 
\begin{equation}\label{sigma*}
\frac{\sigma^* \xi}{K}=\frac{1}{8\exp\left(1-\frac{4E_0}{\pi K}\right)}.
\end{equation}
As previously mentioned, $E_0\sim A_0\xi^2$, whereas the core size $\xi\sim\sqrt{K/A_0}$~\cite{notexi}, so the right hand side of Eq.~\ref{sigma*} is a constant. Therefore within our mean field theory the physics is determined by the dimensionless parameter $\chi\equiv \sigma \xi/K$, controlling the ratio between interfacial and elastic properties -- we call this an inverse elastic capillary number. Eq.~\ref{sigma*} leads to the expectation of a discontinuous transition between a regime where the isotropic component forms cylinders wetting the disclination network (for $\sigma<\sigma^*$) and another regime where such cylinders disappear (for $\sigma>\sigma^*$). Anchoring with finite strength $W$ (units N/m) would introduce another dimensionless parameter, $w=W/\sigma$; reasoning as in~\cite{Voloschenko2002} suggests that anchoring effects are unimportant if $Wr^*/K<1$ ($w<8$). {{Residual effective anchoring due to implicit couplings between ${\bf Q}$ and $\nabla\phi$ is estimated to be small from inspection of the director field profiles (Suppl.~Movie 2).}} 

Thermodynamically, the cubic fluid cylinder phase should be stable when the LC:isotropic ratio is $\sim p/r^*$ but coexist with an excess of one or other pure phase otherwise. Indeed, in Figure~\ref{fig2}(a-e) (and Suppl.~Movie~3) we show that the dynamics for an 85:15 emulsion with $\chi\approx 0.019$ is significantly different from that of Figure~\ref{fig:BPIstruct} (for a 90:10). A liquid tube network still forms early on, but is unstable: the structure later on twists and rearranges leading to the formation of isotropic fluid domains of irregular shape (Fig.~\ref{fig2}(b-d)). These slowly coarsen until they reach a seemingly arrested ``elastic emulsion'' state at late times (Fig.~\ref{fig2}(e)). 

When $\chi$ is increased ($\chi\approx 1.9$, Fig.~\ref{fig2}(f-j) and Suppl. Movie 4, for an 80:20 mixture) the mixture morphology changes again. Following a deep quench into the demixed phase, fluid droplets coarsen and interact with the BPI disclination network, forming an emulsion of spherical droplets (Fig.~\ref{fig2}(f) and {{Suppl.~Movies~4~and~5}}). Droplets are connected to each other by disclinations, creating elastic interactions between them. Nearby droplets can coalesce (see highlighted dotted region in Fig.~\ref{fig2}(g-i)) and small subcritical droplets shrink in favour of larger nearby ones, through a process akin to Ostwald ripening (see highlighted dashed region in Fig.~\ref{fig2}(f,g)). At late times, the dynamics slows down until we again find a {{seemingly}} arrested metastable structure (Fig.~\ref{fig2}(j), and Suppl.~Movie~4). The arrest of Ostwald ripening is expected whenever the continuous phase has a yield stress~\cite{elasticostwald}; this may apply here despite the rewiring of BPI into an amorphous disclination network~\cite{oliverrheology}.
For this value of $\chi$ the droplet-disrupted defect network is reminiscent of structures observed with hard colloids in BPs~\cite{miha1,miha2,natcom}. However, the dynamics of formation of these structures is very different in the two cases. Similar dynamics and arrest as shown in Figure~\ref{fig2}(f-j) is observed for different compositions, including a 90:10 emulsion (Suppl.~Movie~6). 

Experimentally, $\sigma$, $\xi$ and $K$ should be  $\sim$ 1 mN/m, $\sim$ 10 nm and $\sim 10$ pN respectively, yielding a typical $\chi\sim 1$, corresponding to arrested droplets (Fig.~\ref{fig2}(f-j), see also~\cite{Mapping}). It might though be possible to decrease $\chi$ to reach the cubic fluid cylinder regime, for instance by employing as the second component a mesogenic liquid that is chemically similar to the chiral LC but has a lower clearing temperature so that it remains isotropic within the thermal stability range of the BP. The latter could give a $50$-fold reduction in $\sigma$~\cite{langevin}. 

\begin{figure}
\includegraphics[width=\columnwidth]{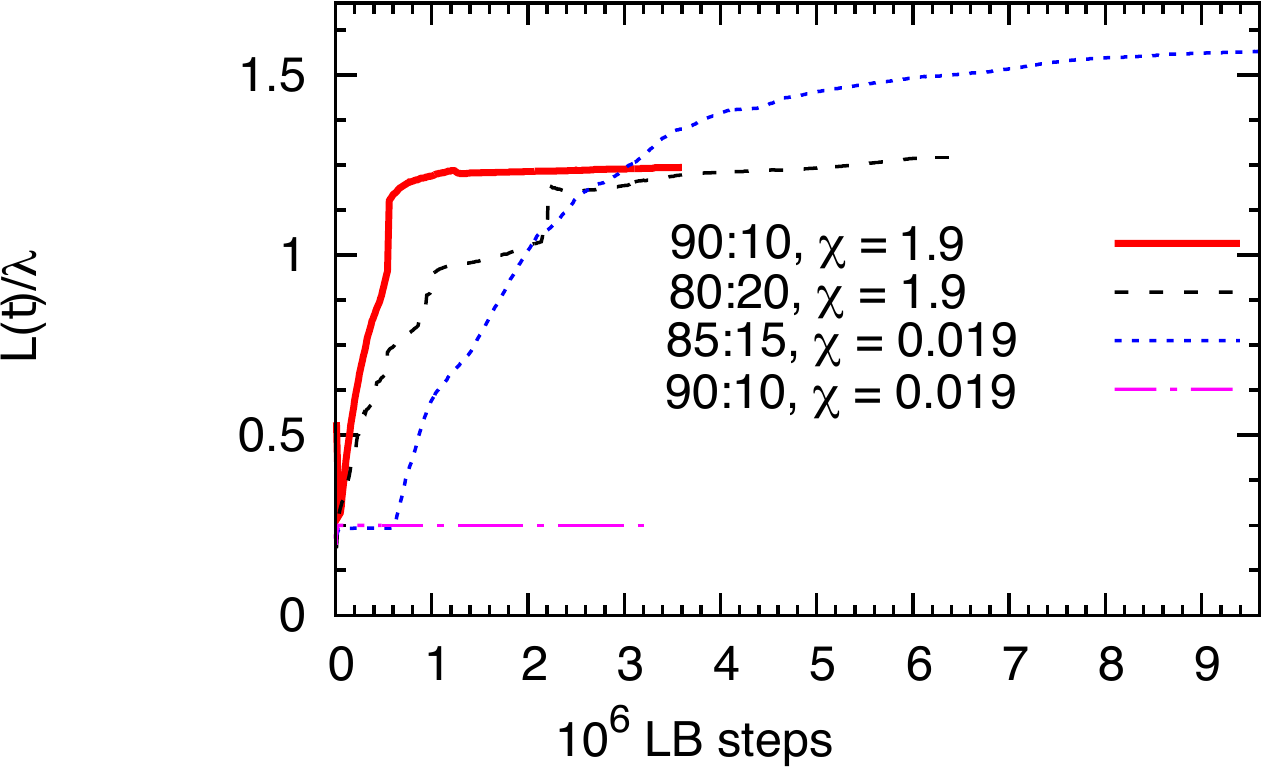}
\caption{Time evolution of the domain length scale $L(t)$ for different values of $\chi$ and emulsion compositions.} 
\label{fig3}
\end{figure}

\begin{figure}
  \includegraphics[width=\columnwidth]{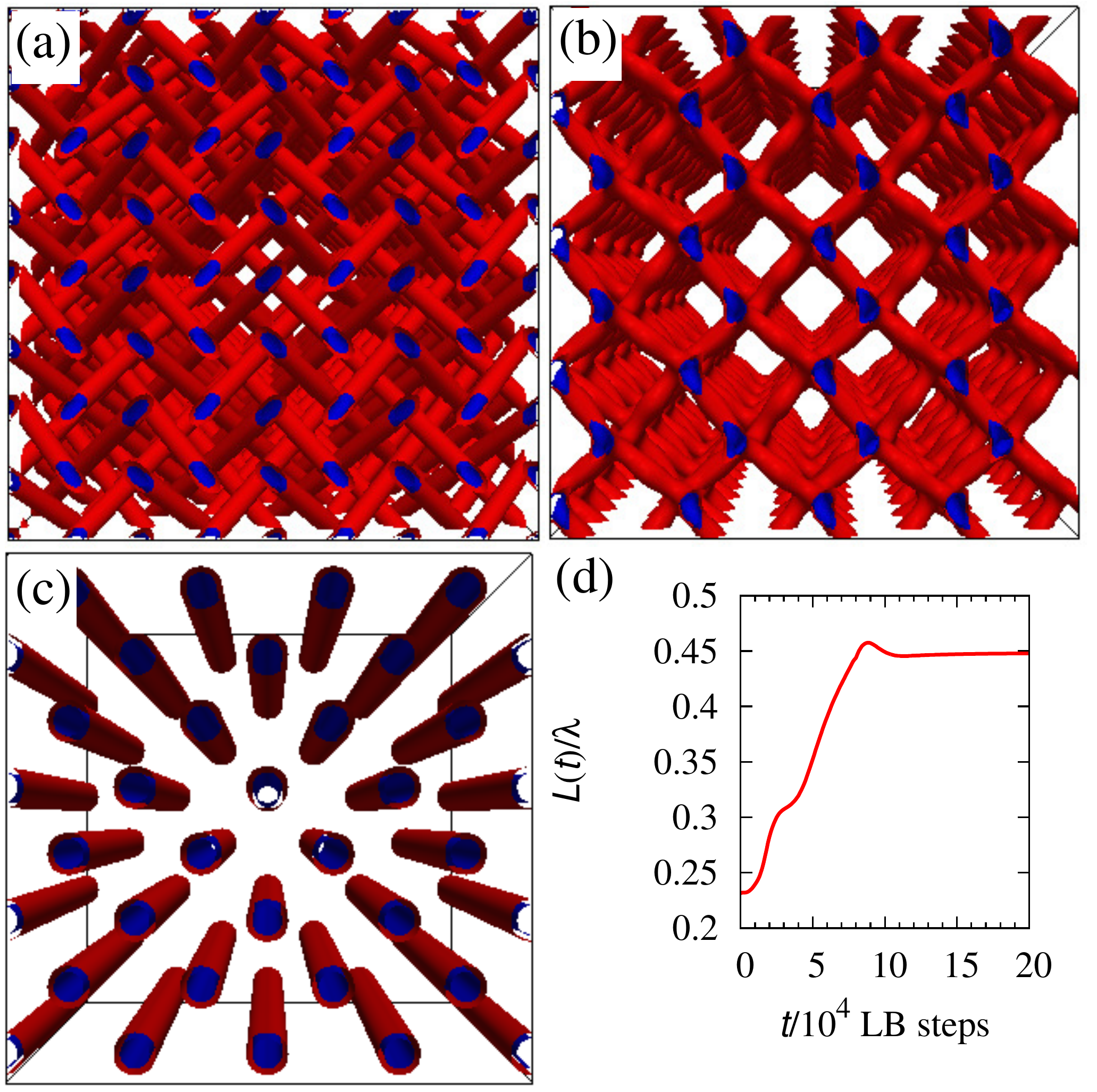}
\caption{Dynamics of a 90:10 mixture of low $\sigma$ ($\chi\approx 0.019$) in a strong electric field along the $z$ axis (into the page). The strength of the field is quantified by the dimensionless quantity $e=\sqrt{27\epsilon_a E^2/(32\pi A_0\gamma)}$: in our case $e\approx 0.24$~(given the mapping in~\cite{Mapping}, this may be realised with $E\sim 25$ V/${\rm \mu}$m  for a LC with $\epsilon_a\sim 1.8\times 10^{-10}$ F/m). (a)~Snapshot of the starting configuration at $t=0$ (no field). When the field is on, the isotropic fluid (red) follows the rapidly reorganising defect network (blue). (b,c)~Snapshots corresponding to $t=5\times 10^{4}$ (b), and to $t=2\times 10^5$ (c). (d)~Plots of the domain length-scale $L(t)/\lambda$ as a function of time. {{(Simulations performed using parameter set {\it D} (see Appendix).)}} 
}
\label{fig4}
\end{figure}

We next consider the time-dependent domain size $L(t) = 2\pi{\int S(k, t)\mathrm{d} \mathbf{k}}/{\int kS(k,t)\mathrm{d} \mathbf{k}}$, where $k=|\mathbf{k}|$~\cite{kendon}. Here $S(k,t)=\langle \phi({\mathbf{k}},t)\phi({\mathbf{-k}}, t)\rangle$ is the structure factor at time $t$. We compare $L$ to the BP unit cell size, $\lambda$.
Figure~\ref{fig3} shows plots of $L(t)$ for all cases presented hitherto. For low $\chi$, corresponding to self-assembly of the cubic fluid cylinder phase (Fig.~\ref{fig:BPIstruct}(b)), $L(t)$ rapidly settles to $L\sim \tfrac{\lambda}{4}$ (for a 90:10 mixture with $\chi\approx 0.019$, dot-dashed line in Fig.~\ref{fig3}).
When the amount of isotropic fluid is increased, the dynamics of $L(t)$ reflects the events visualised in Figure~\ref{fig2}(a-e) (or Suppl.~Movie~2). Initially $L(t)\sim \tfrac{\lambda}{4}$ as a cubic fluid forms. This structure is only metastable due to the excess of isotropic fluid, so later on $L(t)$ shows a rapid growth which appears to saturate at late times. While the curve does not become completely flat in Figure~\ref{fig3}, the dynamics is many orders of magnitude slower than observed with binary simple fluids~\cite{kendon}. 
For the case favouring spherical droplets (Fig.~\ref{fig2}(f-j), and Suppl.~Movie~3), the domain size shows an immediate rapid growth corresponding to coarsening. Later on, $L(t)$ settles to a value $\sim 1.2\lambda$ (solid and dashed lines in Fig.~\ref{fig3}).

We now demonstrate that our BP-based elastic emulsions can be manipulated and controlled with an electric field. 
In Figure~\ref{fig4} (and Suppl.~Movie~5), we show the dynamics of a 90:10 emulsion with low $\chi\approx 0.019$ under a strong electric field along the $z$ axis. Without field, the mixture self-assembles into a cubic fluid network. Under the applied field the isotropic fluid follows closely the reorganising disclination network (Fig.~\ref{fig4}(a-c)), including an intermediate state with square lattice of disclinations perpendicular to the field (Fig.~\ref{fig4}(b)). The time development of $L(t)$ confirms this (Fig.~\ref{fig4}(d)), growing from its $\sim\tfrac{\lambda}{4}$ value in the zero field configuration, to $L\sim 0.45\lambda$: the plot displays two blips corresponding to the formation of the intermediate and final cubic fluids (Figs.~\ref{fig4}(b) and (c), respectively). 

{\it Conclusions:} We have shown that a mixture of a blue phase (BPI) and an isotropic fluid leads to a range of elastic emulsions with fascinating dynamical and phase behaviours. We have identified a key control parameter, $\chi=\sigma \xi/K$, which determines the relative importance of interfacial and elastic forces, and which, together with the mixture composition, determines most of the observed physics. For small values of $\chi$, and sufficiently small concentration of the isotropic component, we observe the self-assembly of a cubic fluid cylinder phase, where the isotropic fluid arranges into cylinders which replace the disclination lines of the BPI network. Unusually, this structure contains unbranched, straight fluid cylinders extending in all three spatial directions with potential applications in materials templating~\cite{ninham}. If the fraction of isotropic component is increased, this cubic fluid structure still forms initially, but is subsequently unstable and collapses to yield irregular isotropic domains connected by disclination lines. For larger values of $\chi$ the isotropic fluid arranges into droplets, which slowly coarsen via a combination of coalescence and an Ostwald-type process. At very late times, we find an apparently arrested metastable structure whose typical size is comparable with the BP unit cell. The cubic fluid cylinder phase found at low $\chi$ can additionally be manipulated with an external field, and we have shown that in this way we can switch between cubic fluids with different symmetries and/or network topologies. 

{\it Acknowledgements:} This work was funded by EU intra-European fellowship 623637 DyCoCoS FP7-PEOPLE-2013-IEF, IdEx Bordeaux junior chair to JSL and UK EPSRC grant EP/J007404/1. Part of this work was performed using HPC resources from GENCI-IDRIS (Grant 2017-20907383). MEC is funded by the Royal Society.

\begin{appendices}

\section{Appendix}

\subsection{Simulation parameters and thermodynamic stability of blue phase I}

To simulate the blue phase I (BPI) and isotropic fluid mixture, we consider a total free energy ${\cal F} = \int fdV$. The free energy density $f$ consists of two parts, $f=f_\phi + f_Q$, where the $f_\phi$ describes a binary fluid and $f_Q$ describes the liquid crystal (LC) ordering. These are given in Eqs. (1-2) in the main text.
\\

The binary fluid part (Eq. (1) in the main text) depends on three positive constants: $a, b$ and $\kappa$, which can be used to define an interfacial width $\xi_{\phi}=\sqrt{2\kappa/a}$ and an interfacial tension $\sigma = \sqrt{8\kappa a^3/(9b^2)}$. All simulations use the same value of $\xi_{\phi}\approx 1.3$, while the interfacial tension $\sigma$ was varied (see below for details). Additionally, $a=b$ was chosen, so that $\phi = \pm 1$ corresponds to a fully orderded LC phase, and an isotropic fluid, respectively.
\\

For the chiral liquid crystal component, the free energy functional is expressed in terms of an order parameter tensor $\mathbf{Q}$ and given by Eq. (2) in the main text. In the equation 2, $A_0$ fixes the energy scale, $K$ is the elastic constant penalising distortions and $q_0=2\pi/p$ sets the equilibrium pitch length $p$ of the chiral liquid crystal. $\gamma(\phi)$ is a temperature like parameter, which controls the order
\begin{equation}
  \gamma(\phi) = \gamma_0 +\delta(1+\phi).
\end{equation}
The parameters $\gamma_0 = 2.586$ and $\delta = 0.25$ were chosen such that for the isotropic phase $\gamma(-1) = 2.586$ is below the order-disorder transition at $\gamma_c\sim 2.7$, while for the ordered phase $\gamma(+1) = 3.086$. The thermodynamic state of the chiral LC is determined by the reduced chirality
\begin{equation}
  \kappa_{\mathrm{LC}}=\sqrt{108Kq^2_0/(A_0\gamma})
\end{equation}
and the reduced temperature
\begin{equation}
  \tau = 27\left(1-\gamma/3 \right)/\gamma.
\end{equation}
The same values of $\kappa$ and $\tau$ was used for all simulations. Using the values for the ordered phase ($\phi= + 1$), these are $\kappa_{\mathrm{LC}}\approx 0.7$ and $\tau \approx 0.25$, giving a thermodynamically stable BPI~\cite{OliBP}. We also define a typical defect core size as $\xi=\sqrt{K/A_0}$ -- this is analogous to the nematic correlation length defined in~\cite{mermin} but misses out the dimensionless prefactor (which in our case would depend on $\gamma$, hence on $\phi$).
\\

All the simulations were carried out in cubic simulations box spanning $128\times 128\times 128$ lattice points, with periodic boundary conditions. Throughout the simulations the fluid viscosity was set to $\eta = 0.5$ and the rotational viscosity of the LC  $\Gamma = 0.5$ (in Eq. (3) of the main text).\\

In Eq. (3), the ${\bf S}\left({\bf W},{\bf Q}\right)$ term, accounting for flow-induced LC rotations, is explicitly given by~\cite{LBLC}
\begin{eqnarray}
\nonumber 
\mathbf{S}(\mathbf{u},\mathbf{Q}) & = &  \left( \nu \mathbf{D} + \mathbf{\Omega} \right) \left( \mathbf{Q} + \mathbf{I}/3 \right) 
+ \left( \mathbf{Q} + \mathbf{I}/3 \right) \left( \nu \mathbf{D} 
-  \mathbf{\Omega} \right) \\ 
& - & 2 \nu \left( \mathbf{Q} + \mathbf{I}/3 \right) Tr \left(\mathbf{QW} \right),
\label{Stensor}
\end{eqnarray}
where $\mathbf{D} = ( \mathbf{W} + \mathbf{W}^T ) / 2$ and $\mathbf{\Omega} = ( \mathbf{W} - \mathbf{W}^T ) / 2$ are the symmetric and anti-symmetric parts of the velocity gradient tensor $W_{\alpha \beta} = \partial_{\beta} u_{\alpha}$ respectively, and $\mathbf{I}$ is the identity matrix. The quantity $\nu$ is the flow aligning parameter -- we choose $\nu=0.7$ which corresponds to a flow-aligning LC. \\

The stress tensor of the Navier-Stokes equation, $\Pi_{\alpha\beta}$, has the following explicit form~\cite{LBLC},
\begin{eqnarray}
  \Pi_{\alpha \beta}   & = &  -P_0 \delta_{\alpha \beta} +  2 \nu \left( Q_{\alpha \beta} + \frac{1}{3} \delta_{\alpha \beta} \right) Q_{\gamma \mu} H_{\gamma \mu} 
\\ \nonumber & - & 
- \nu H_{\alpha \gamma} \left( Q_{\gamma \beta} + \frac{1}{3} \delta_{\gamma \beta} \right) -  \nu \left( Q_{\alpha \gamma} + \frac{1}{3} \delta_{\alpha \gamma} \right) H_{\gamma \beta} \\ \nonumber
& - &\frac{\partial f }{\partial \partial_{\beta} Q_{\gamma \mu} }\partial_{\alpha} Q_{\gamma \mu} + Q_{\alpha \nu} H_{\nu \beta} - H_{\alpha \nu} Q_{\nu \beta},
\end{eqnarray}
where Greek indices denote Cartesian coordinates and summation over repeated indices is implied.

Regarding the rest of the parameters: $a=b, \kappa$ and mobility $M$ for the binary fluid as well as $A_0, K$ and $q_0$ (as well as the field $E$ when present) for the LC, were varied, so that the conditions of constant interfacial width, and the thermodynamic stability of BPI were retained (as explained above) -- values for each of the Figures are given below (key dimensionless parameters are also reported in the Figure captions in the main text. \\

For all simulations reported, the the ${\bf Q}$ was initialised with the high-chirality limit solution corresponding to BPI (see~\cite{OliBP} for details). This is appropriate as the values of $\kappa_{\mathrm{LC}}$ and $\tau$ are such that BPI is the thermodynamically stable phase for the liquid crystal component. In Figures 2(a-e) and 4, the compositional order parameter was initially set to $\phi_0=(A-B)/(A+B)$ with small noise, where $A$ and $B$ denote the fraction of LC and isotropic phase respectively. This corresponds to an initially demixed configuration. In Figures 1 and 2(f-j), the isotropic fluid was initialised into tiny droplets to trigger phase separation (in view of the large asymmetry between fractions of $A$ and $B$). Such an initialisation may also  be used in Figures 2(a-e) and 4, yielding analogous results as shown in the main text. 

\subsection{Simulation parameters for the Figures in the main text}

{\bf Figure 1} in the main text uses parameter set {\it A}: $K = 0.14$, $A_0 = 0.05$, lattice parameter for the BPI $\lambda = p/\sqrt{2} = \sqrt{2}\pi/q_0 = 64$ for the LC and $a=b=6.25\times 10^{-4}$, $\kappa \approx 5.3\times 10^{-4}$ and $M = 5$ for the binary fluid. These correspond to an inverse capillary number $\chi = \sigma\xi/K\approx 0.019$, where $\xi=\sqrt{\frac{27 K}{A_0 \gamma}}\simeq 4.9$ (see main text for details on the physical meaning of $\chi$, note the relevant value of $\gamma$ is that in the ordered phase, at $\phi=+1$). We have verified that qualitatively indistinguistable results are obtained when including a ``redshift'' $r=0.83$ for BPI, so that $\lambda=p/(r\sqrt{2})$ -- which leads to a lower free energy in single-component BPI~\cite{OliBP}. \\

{\bf Figure 2}(a-e) in the main text uses parameter set {\it B}: $K = 0.07$, $A_0 = 0.1$, lattice parameter for the BPI $\lambda = 32$ for the LC and $a=b=6.25\times 10^{-4}$, $\kappa \approx 5.3\times 10^{-4}$ and $M = 0.5$ for the binary fluid. These give $\xi\sim 2.5$ and $\chi \approx 0.019$. Note that as our definition of $\xi$ does not include dimensionless prefactors,  a defect core may in practice still be resolved by a few lattice sites in our simulations even when $\xi \sim 1$ (see Suppl. Movies 2 and 4).\\

{\bf Figure 2}(f-j) in the main text uses parameter set {\it C}: $K = 0.14$, $A_0 = 0.05$, lattice parameter for the BPI $\lambda = 64$ for the LC and $a=b=6.25\times 10^{-2}$, $\kappa \approx 5.3\times 10^{-2}$ and $M = 0.05$ for the binary fluid. These give $\xi \sim 4.9$ and $\chi \approx 1.9$. \\

{\bf Figure 4} in the main text uses parameter set {\it D}: $K = 0.07$, $A_0 = 0.1$, lattice parameter for the BPI $\lambda = 32$  for the LC and $a=b=6.25\times 10^{-4}$, $\kappa \approx 5.3\times 10^{-4}$ and $M = 0.5$ for the binary fluid. 
These give $\xi \sim 2.5$ and $\chi \approx 0.019$. 
The electric field is $\mathbf{E}=[0, 0, 0.04]$, and the dielectric anisotropy is $\epsilon_a=41.4$. \\

\subsection{Mapping to SI units}

Assuming typical values for chiral liquid crystals $K\sim 10$pN and $\lambda \sim 300$nm, we can map an interfacial tension value of $\sigma = 1$ in simulations to $\sigma=20$ mN/m in physical units. This gives a typical interfacial tension $\sigma\approx 10^{-5}$ N/m for the cubic fluid (Figure 1 in the main text), while the elastic emulsion (Figure 2(f-j)) is observed when $\sigma \approx 10^{-3}$N/m.

\subsection{Captions for Supplementary Movies}

{\bf Suppl. Movie 1:} Movie of the dynamics corresponding to Figure 1 in the main text. \\

{\bf Suppl. Movie 2:} Slices of the director profile through the system (planes perpendicular to the $z$ axis), for the final configuration in Figure 1. The director field is defined for each non-zero value of the ${\bf Q}$ tensor, but the associated order is very weak within the isotropic phase (blue). It can be seen that the effective anchoring if present is small, as there is not a well-defined angle between the director field and the surface normal.
\\

{\bf Suppl. Movie 3:} Movie of the dynamics corresponding to Figure 2(a-e) in the main text.
\\

{\bf Suppl. Movie 4:} Movie of the dynamics corresponding to Figure 2(f-j) in the main text.
\\

{\bf Suppl. Movie 5:} Slices of the director profile through the system (planes perpendicular to the $z$ axis), for the configuration in Figure 2j. Similar considerations apply as for Suppl. Movie 2 -- the effective anchoring though appears normal in this case.
\\

{\bf Suppl. Movie 6:} Movie of the dynamics corresponding to a system with the same parameters as in Figure 2(f-j) (i.e., parameter set {\it C}), but with 90:10 composition.
\\

{\bf Suppl. Movie 7:} Movie of the dynamics corresponding to Figure 4 in the main text.
\\

\end{appendices}

\end{document}